\documentclass[twocolumn,prb,showpacs,floatfix,superscriptaddress]{revtex4-1}
\usepackage{graphicx}
\usepackage{dcolumn}
\usepackage{bm}
\usepackage{amssymb}
\usepackage{amsmath}
\usepackage{amsfonts}
\usepackage{subfigure}
\usepackage{hyperref}
\usepackage{xcolor}%
\providecommand{\U}[1]{\protect\rule{.1in}{.1in}}
\newcommand{\Gd}{GdPtPb}

\providecommand{\U}[1]{\protect\rule{.1in}{.1in}}
\begin{document}
\title{GdPtPb: A  non collinear antiferromagnet with distorted Kagom\'e lattice}
\author{S. Manni}
\affiliation{ Department of Physics and Astronomy, Iowa State University, Ames, IA 50011, USA }
\affiliation{Ames Laboratory, Iowa State University, Ames, IA 50011, USA }
\author{Sergey L. Bud'ko}
\affiliation{ Department of Physics and Astronomy, Iowa State University, Ames, IA 50011, USA }
\affiliation{Ames Laboratory, Iowa State University, Ames, IA 50011, USA }
\author{Paul C. Canfield}
\affiliation{ Department of Physics and Astronomy, Iowa State University, Ames, IA 50011, USA }
\affiliation{Ames Laboratory, Iowa State University, Ames, IA 50011, USA }

\date{\today}

\begin{abstract}
	In the spirit of searching for Gd-based, frustrated, rare earth magnets, we have found antiferomagnetism (AF) in GdPtPb which crystallizes in the ZrNiAl-type structure that has a distorted Kagom\'e lattice of Gd-triangles. Single crystals were grown and investigated using structural, magnetic, transport and thermodynamic measurements. GdPtPb orders antiferromagnetically at 15.5 K arguably with a planar, non-collinear structure. The high temperature magnetic susceptibility data reveal an "anti-frustration" behavior having a frustration parameter, $|f|$ = $|\Theta|$/ $T_N$ = 0.25, which can be explained by mean field theory (MFT) within a two sub-lattice model. Study of the magnetic phase diagram down to $T$ = 1.8 K reveals a change of magnetic structure through a metamagnetic transition at around 20 kOe and the disappearance of the AF ordering near 140 kOe. In total, our work indicates that, GdPtPb can serve as an example of a planar, non collinear, AF with a distorted Kagom\'e magnetic sub-lattice.       
	
\end{abstract}

\maketitle


\section{Introduction}
 
Magnetic frustration in insulators can lead to intriguing ground states such as quantum spin liquids (QSL)~\cite{Balnets, Kitaev} or spin ices.~\cite{ice} Magnetic frustration is usually realized in geometrically frustrated  pyrochlore, triangular, Kagom\'e or hyperkagom\'e spin sub-lattices with a localized, often nearest neighbor, description of magnetic spin exchanges.\cite{Ramirez} This approach brings a fundamental difficulty to the description of magnetic frustration in intermetallic systems which have longer range spin-spin interactions.

 Recently, significant attempts were made to realize and theoretically understand magnetically frustrated ground states in geometrically frustrated rare earth intermetallic compounds.\cite{Ballaou, Coleman,CePdAl1,CePdAl2,CeRhSn,YbAgGe1,YbAgGe,Yb2Pt2Pb,SSL,tmb4} The main focus of these efforts has been concentrated around either a quasi-Kagom\'e lattice with ZrNiAl-type structure or a Shastry-Sutherland lattice with the U$_2$Pt$_2$Si-type structure.\cite{CePdAl1,CePdAl2,CeRhSn,YbAgGe1,YbAgGe,Yb2Pt2Pb,SSL} Due to the long range nature of the RKKY-interaction, realizing a QSL state seems to be a difficult goal to achieve in intermetallic compounds where, in general, a magnetically ordered ground state is achieved by the longer range magnetic exchange  and/or with the help of quantum disorder or lattice disorder.\cite{Ballaou} Rather intermetallic compounds offer a rich variety of magnetic ground states both as a function of temperature as well as a function of applied magnetic field. Examples include CePdAl and YbAgGe both with the ZrNiAl-type structure, and Yb$_2$Pt$_2$Pb with the U$_2$Pt$_2$Si-type structure.\cite{CePdAl1,CePdAl2,YbAgGe1,YbAgGe,Yb2Pt2Pb} On the other hand there are some promising (and debated) examples of potential metallic spin liquids such as CeRhSn~\cite{CeRhSn} and Pr$_2$Ir$_2$O$_7$~\cite{Pr2Ir2O7} which are paramagnetic down to lowest temperature despite strong AF spin correlations. 
 
We have focused our search for magnetically frustrated ground states in rare earth intermetallic systems with the ZrNiAl-type structure. In this structure, rare earth ions form a distorted Kagom\'e lattice in the $ab$-plane and are stacked along the  $c$-axis. If the interlayer distance of the $ab$-planes is much larger than rare earth distances in the $ab$-plane, the possibility of low dimensional frustrated exchange interaction arises. In the $R$PtPb ($R$ = rare earth ion) intermetallic series, CePtPb was reported~\cite{CePtPb} to be an antiferromagnet with low $T_N$ = 0.9 K, similar to YbAgGe. In many Ce and Yb-based, frustrated intermetallics magnetic exchange is governed by ground state doublets ($J$ = 1/2). Often, due to  crystal electric field splitting, magnetic anisotropy influences the magnetic exchange interaction. This can be the case for all rare earths except Gd$^{3+}$ and Eu$^{2+}$-based ones. Hence, we wanted to explore a Gd-based, geometrically frustrated lattice where, due to absence of crystal electric field effect, the whole $J$ = 7/2 multiplet participates in magnetic exchange interaction and the Gd$^{3+}$ has $J$ = $S$ = 7/2 Heisenberg moment. 

 We have grown and studied single crystals of GdPtPb, which crystallizes in the same crystal structure as CePtPb and synthesized in single crystalline form. GdPtPb orders antiferromagnetically  below 15.5 K and, most interestingly, it shows "anti-frustration" behavior having a frustration parameter $|f| = |\Theta| / T_N$ much less than one. Magnetic susceptibility suggests a possible special, non-collinear antiferromagnetic, structure. Overall we have characterized GdPtPb structurally, magnetically and thermodynamically and tried to relate its magnetism to its underlying, at first glance-geometrically frustrated, magnetic sub-lattice.                       

\section{Experimental and structural details}
\begin{figure}
	\centering
	\includegraphics[width=\columnwidth]{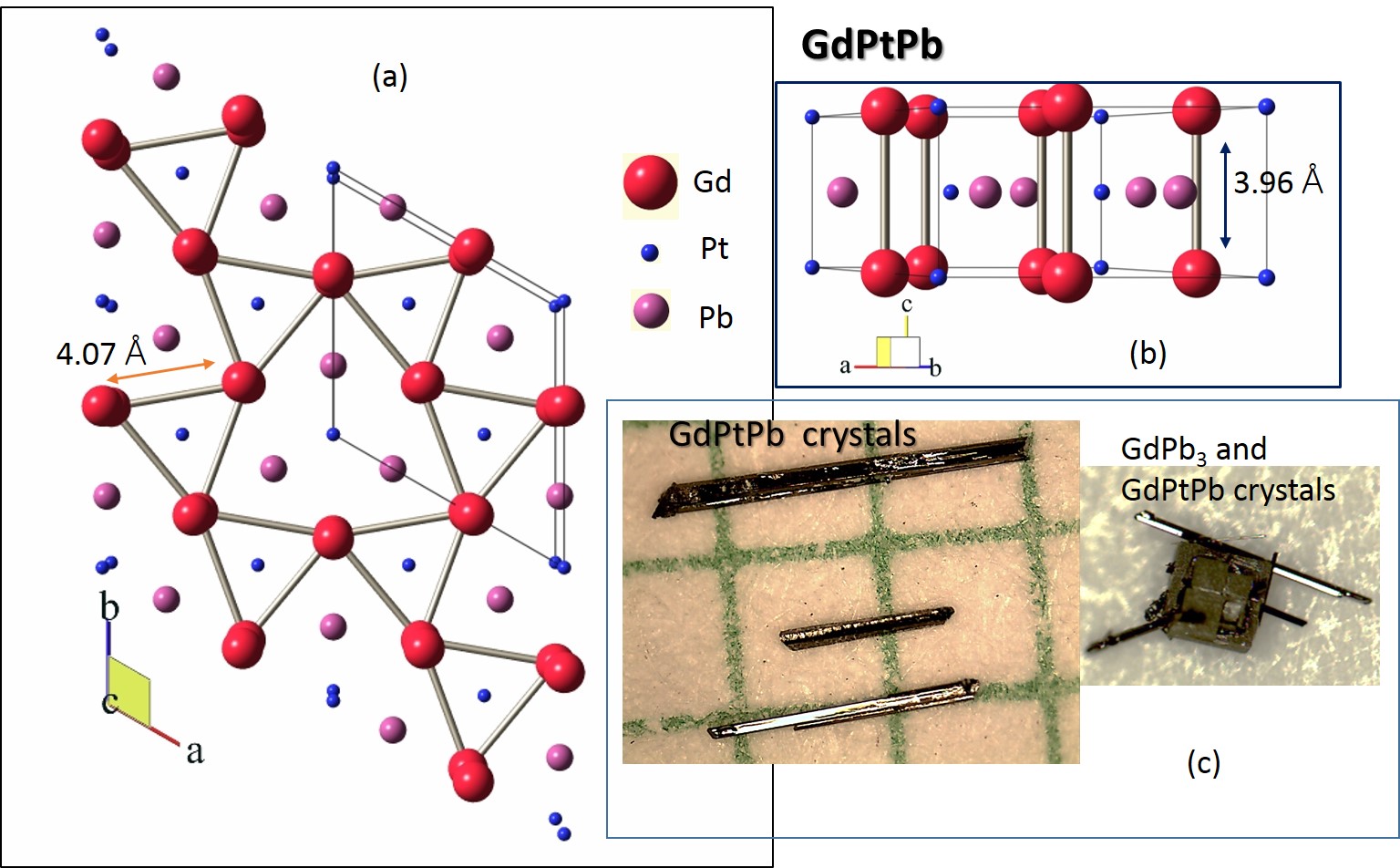}
	\caption{(a) Crystal structure of GdPtPb shown in $ab$- plane. It shows a distorted Kagom\'e lattice of the Gd triangles. Red balls represent Gd; blue Pt and pink Pb, the size of the balls are not according the scale of atomic radius. (b) Crystal structure perpendicular to $ab$-plane. (c) Hexagonal, rod-like GdPtPb crystals on a mm grid and growth of GdPtPb rods on a GdPb$_3$ cubic crystal.}\label{xtal}
	
\end{figure}
\Gd~single crystals are grown from a Pb-rich solution with an initial stoichiometry of Gd:Pt:Pb = 5:5:90. Elemental, pure ($\ge$99 \%), metals were packed in 2 ml fritted Al$_2$O$_3$ crucible set and then sealed in a quartz ampule under partial pressure of Argon before putting in furnace.\cite{canfield} The whole assembly was heated to 1180$^{\circ}$C and cooled down to 600 $^{\circ}$C at a 5$^{\circ}$C/hour rate after which the remaining, Pb-rich, solution was decanted. We obtained millimeter size, hexagonal, rod-like crystals of GdPtPb (Fig.~\ref{xtal}c) and some GdPb$_3$ impurity phase, often as cubic single crystals, shown in  Fig.~\ref{xtal}(c). In general GdPtPb and GdPb$_3$ were not inter-grown, although when GdPb$_3$ was present it often had some GdPtPb rods attached to it. In a very similar method, LaPtPb, hexagonal rod-like, crystals were grown from a solution with an initial stoichiometry of La:Pt:Pb = 10:10:80. LaPtPb crystals were used to estimate the non-magnetic contribution to the specific heat of the \Gd-system.

\begin{figure}
	\centering
	\includegraphics[width=\columnwidth]{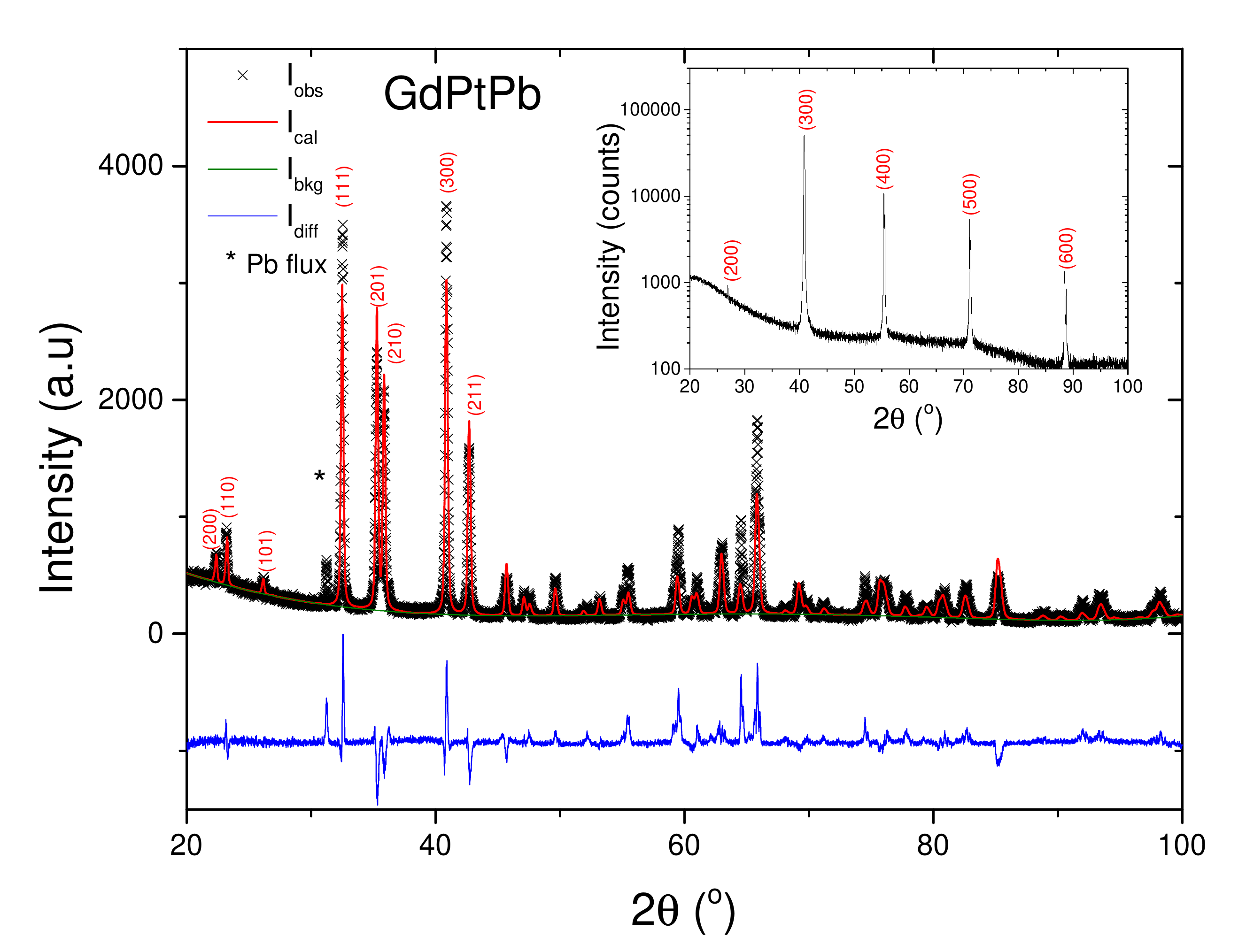}
	\caption{Powder x-ray diffraction  pattern of GdPtPb ground single crystals ($I_{obs}$), Rietveld refinement of the pattern with $P-62/m$ crystal structure ($I_{cal}$) and $I_{diff}$ = $I_{obs}$-$I_{cal}$. Inset shows $\theta$ - 2$\theta$ scan on one GdPtPb single crystal for x-ray incidence angle $\theta$ with the plane perpendicular to rod axis.}\label{xrd}
\end{figure}

\begin{table}[h]
	\caption{Structural details of GdPtPb obtained from Rietveld analysis of powder x-ray diffraction data (see Fig.~\ref{xrd})}%
	\label{abc}%
	\centering
	\begin{tabular}{*3{c}}\hline
		Crystal system & \vline & Hexagonal \\
		Space group & \vline & $P-62m$\\
		a &  \vline &7.637(12) \AA\\
		c & \vline & 3.9649(6) \AA \\
		$\alpha$ & \vline & 90$^{\circ}$\\
		$\beta$ & \vline & 90$^{\circ}$\\
		$\gamma$ & \vline & 120$^{\circ}$\\
		Cell volume & \vline & 200.26(8) \AA$^3$\\
		\hline
	\end{tabular}
\end{table}

\begin{table}[h]
	\caption{Atomic coordinates of the GdPtPb structure obtained from Rietveld analysis of the powder x-ray diffraction data (see Fig.~\ref{xrd})}%
	\label{atom}%
	\centering
	\begin{tabular}{*5{c}}\hline
		Atom & Wyck &  $x$ & $y$ & $z$ \\
					\hline\hline
				
		Pb & 3g & 0.26558(30) & 0.0 & 0.5 \\
		Gd & 3f & 0.6066(5) & 0.0 & 0.0 \\
		Pt & 2d & 0.33333 & 0.66667 & 0.5 \\
		Pt & 1f & 0.0 & 0.0 & 0.0 \\
		\hline
	\end{tabular}
\end{table}
To determine the structure of GdPtPb, powder x-ray diffraction was done on crushed single crystals using a Rigaku Miniflex diffractometer and fitted with published crystal structure of CePtPb by Rietveld refinement method using GSAS-EXPGUI software.\cite{gsas,expgui} CePtPb is reported to be crystallized in hexagonal $P-62/m$ crystal structure.\cite{CIF,CePtPb} Fig.~\ref{xrd} shows measured powder diffraction data ($I_{obs}$), fitting with $P-62m$  crystal structure ($I_{cal}$) and difference between measured data and fitting ($I_{diff}$). We have observed a single Pb-impurity peak which was estimated to correspond to less than 5\% elemental Pb (most likely residual droplets of flux on the surface of the crystals) in the phase. The inferred crystal structure parameters are listed in Table~\ref{abc}. The lattice parameters reported for CePtPb are $a$ = 7.73 \AA~and $c$ = 4.13 \AA~and volume is 213.4 \AA$^3$.\cite{CIF} Comparing these values with the parameters listed in Table~\ref{abc}, we can confirm a lathanide contraction in GdPtPb, compared to CePtPb.\cite{CIF}    

 The GdPtPb crystal structure is drawn from the refine lattice parameters (Table~\ref{abc}) and atomic coordinates(Table~\ref{atom}), shown in Fig.~\ref{xtal}. In the $ab$-plane, Gd triangles form a distorted Kagom\'e network [see Fig.~\ref{xtal} (a)]. In the $ab$-plane, the Gd-Gd distance is 4.07\AA. The Kagom\'e network of Gd triangles  are layered along $c$-axis. The Gd-Gd interlayer distance is 3.96 \AA. In this structure, if we consider the longer range RKKY interaction between Gd$^{3+}$ spins on a frustrated Kagom\'e lattice, we can expect an unconventional magnetically ordered ground state in GdPtPb.  
 
 
 To determine the crystallographic $c$-axis and the $ab$-plane on the hexagonal, rod-like crystals we have done a $\theta$ - 2$\theta$ scan on one piece of single crystal.\cite{Anton} The rod-like crystal is placed on the XRD zero reflection puck in such a way that x-ray beam is incident with $\theta$ angle with respect to the plane perpendicular to the axis of the rod. We obtained only ($h00$) reflections; the inferred  value of the $a$ lattice parameter is 7.65 \AA, which is very close the value listed in Table~\ref{abc}. This confirms that the plane perpendicular to the rod direction is $ab$-plane and $c$-axis is along the rod. 

Magnetic measurements were done using a Quantum Design Magnetic Property Measurement System SQUID magnetometer in the 1.8-300 K temperature range and 0 - 55 kOe magnetic field range. Mostly the measurements were done on single crystal pieces of 0.2 - 1 mg mass. Electrical resistivity was measured by a standard four probe method on a rectangular bar like crystal (dimensions: $A$ $\approx$ 0.01 mm$^2$, $l$ $\approx$ 1.15 mm) in a Quantum Design Physical Property Measurement System using ac transport technique (1 mA excitation current and 17 Hz frequency). The largest dimension of the bar was along $c$-axis and current was applied in this direction. For the heat capacity measurements we used five GdPtPb single crystals with a total mass around 2 mg and aligned them on the heat capacity puck such that the applied field was always within the $ab$-plane. Heat capacity measurements on the LaPbPt were done with similar mass of crystals. Measurements were done in a Quantum Design Physical Property Measurement System by relaxation method in the 1.8 - 60 K temperature range and 0 - 140 kOe field range.    
\section{Results}
\begin{figure}[ht]
\centering
\includegraphics[width=\columnwidth]{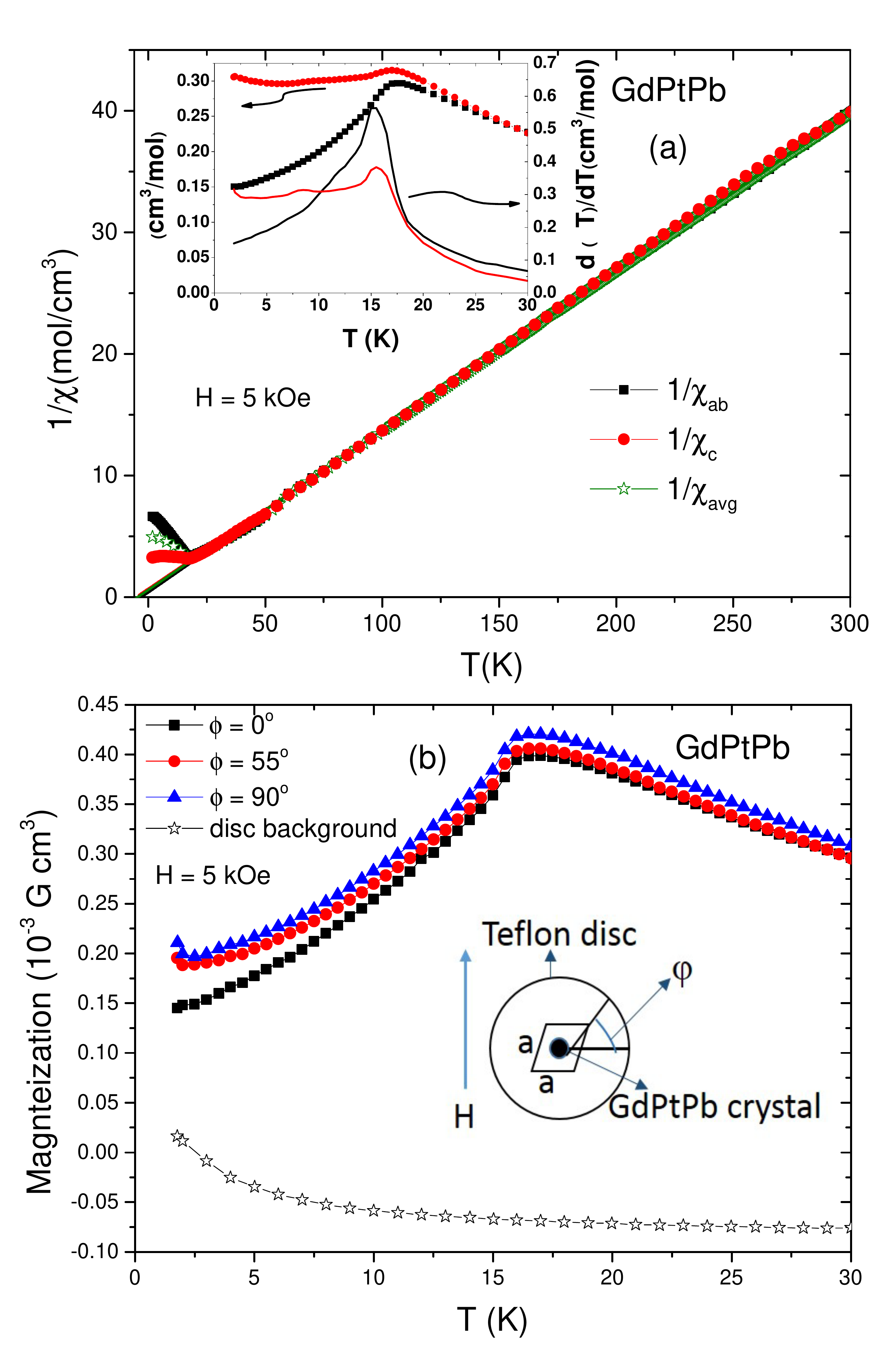}
\caption{(a)Temperature dependent inverse magnetic susceptibility data for $ H||ab$ ($\chi_{ab}$), $H||c$ ($\chi_c$), measured at $H$ = 5 kOe  and average value 1/$\chi_{avg}$ and fitting of the data above 100 K with a Curie Weiss (CW) temperature dependence. Inset shows magnetic susceptibility and $d(\chi T)/dT$ vs. $T$ below 30 K.(b) Temperature dependent magnetization near $T_N$ with three different rotation angle ($\phi$) in $ab$-plane and field along $ab$-plane, measured at $H$ = 5 kOe. The disc background signal is also shown. Inset shows mounting scheme of the rotation measurement and definition of $\phi$.}\label{MT}

\end{figure}

Fig.~\ref{MT}(a) shows the temperature dependent, inverse magnetic susceptibility (1/$\chi$ = $H / M$) plot, measured in a 5 kOe field along the $ab$-plane ($\chi_{ab}$) and the $c$-axis ($\chi_c$) and inverse of calculated average magnetic susceptibility ($\chi_{avg}$) where, $\chi_{avg}$ = (2$\chi_{ab}$+$\chi_{c}$)/3. The inset shows an expanded view of the low-temperature $\chi$($T$) data. $\chi_{ab}$ drops to roughly one half of its maximum value at the lowest measured temperature, and $\chi_{c}$ changes its slope and remains almost constant with a slight low temperature upturn below 5 K (see inset).  The $d(\chi T)/dT$ data (inset, Fig.~\ref{MT}(a)) for both field directions have a maximum at T = 15.5 K (see inset).  This clearly indicates that \Gd~is an antiferromagnet with $T_N$= 15.5 K. High temperature magnetic susceptibility is isotropic and follows the Curie-Weiss (CW) behavior $1 / \chi = (T-\Theta) / C $ where $C$ is the Curie constant reflecting effective moment ($\mu_{eff} \approx \sqrt{8 C}$) and $\Theta$ is the CW temperature reflecting the average magnetic exchange interaction. The fitting is done over different temperature ranges: 60 K to 300 K; 75 K to 300 K; 100 K to 300 K and 150 K to 300 K, variation of the fitted parameters are indicated within the parenthesis. By fitting inverse $\chi_{ab}$, $\chi_c$ and $\chi_{avg}$, we have obtained $\mu_{eff}$ = 7.82 ($\pm$0.01) $\mu_B$, 7.74 ($\pm$0.01) $\mu_B$ and 7.8 ($\pm$0.01) $\mu_B$ respectively, very close to theoretical value of Gd$^{3+}$ (7.94 $\mu_B$) and the CW temperature ($\Theta$)$_{ab}$ = -5.12 ($\pm 1$) K, ($\Theta$)$_{c}$ = -2.78 ($\pm 1$) K and ($\Theta$)$_{avg}$ = -4.2 ($\pm 1$) K respectively. Notably,  $|\Theta_{ab}|$, $|\Theta_{c}|$  $<<$ $T_N$ which is discussed in the context of the mean field theory below. 

The low-$T$ magnetic susceptibility along the easy-plane ($ab$-plane) extrapolates to a finite value at $T$ = 0 K, which hints that magnetic structure  is a non-collinear AF type.  We obtained  $M_{ab}$($T$ = 1.8 K)/ $M_{ab}$($T_N$)  =  0.43 - 0.51 in multiple measurements. To prove that it is a robust effect, we have measured temperature dependent magnetization ($M$) by rotating the crystal in the $ab$-plane and applying field along $ab$-plane. For this measurement we mounted the rod-like crystal inside a teflon disc, at the center of the disc making the rod perpendicular to the disc surface. Hence the $ab$-plane is parallel to the disc plane (shown in inset of Fig.~\ref{MT} (b)). For the different rotation angles the disc is rotated keeping it vertical in the straw such that the applied field is always parallel to the disc plane. The rotation angle is measured with respect to a mark on the teflon disc which has an arbitary angle with the $a$-axis (inset Fig.~\ref{MT} (b)) For the three rotation angles $\phi$ = 0 ($\pm$5)$^{\circ}$, 55 ($\pm$5)$^{\circ}$ and 90($\pm$5)$^{\circ}$, $M_{ab}$($T$ = 1.8 K)/ $M_{ab}$($T_N$)  =  0.37, 0.46 and 0.47 respectively, shown in Fig.~\ref{MT} (b). Our multiple $H || ab$ measurements lead us to conclude that (i) there is little in-plane anisotropy and (ii) $\chi_{ab}$($T$ $\to$ 0)/ $\chi_{ab}$($T_N$) $\approx$  1/2.

\begin{figure}
	\centering
	\includegraphics[width=\columnwidth]{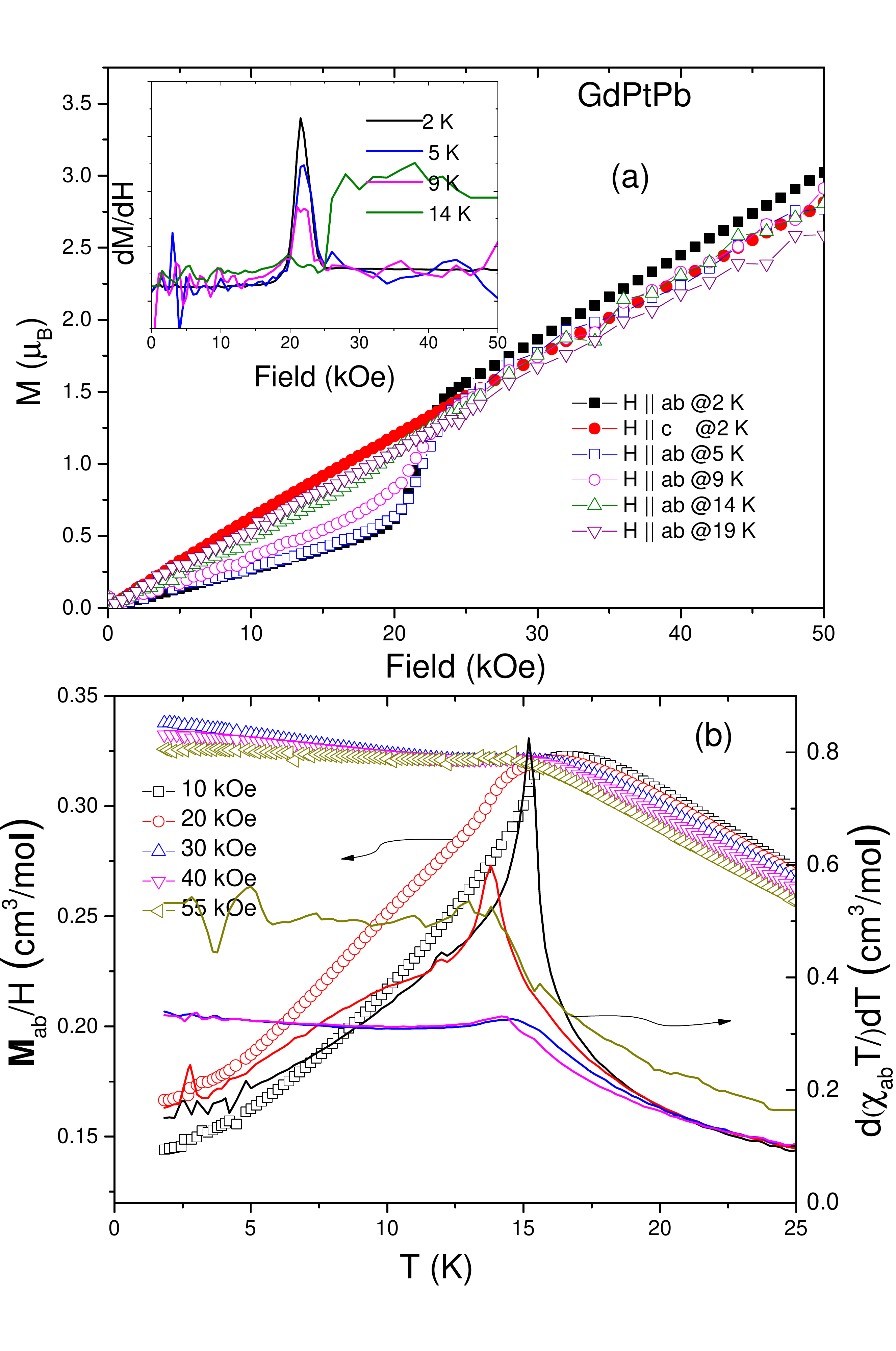}
	\caption{(a)Field dependent magnetization measurement at different temperature and field for $H||ab$ and $H||c$ and inset shows $dM/dH$ vs. $H$ for $H || ab$. (b)The temperature dependent $M_{ab} / H$ and $d(\chi_{ab} T)/dT$ near $T_N$ for different applied field values.}\label{MH}
	
\end{figure}

Figure~\ref{MH} (a) shows magnetization isotherms at different temperatures for $H || ab$ ($M_{ab}$) and at T= 2 K for $H||c$ ($M_c$). A sharp metamagnetic transition is evident in $M_{ab}$ and $dM/dH$ at 22 kOe, which broadens with increasing temperature, and vanishes above the $T_N$. $M_c$ is proportional to field having no evident metamagnetic transition in the measured field range.   Below 22 kOe, a clear anisotropy exists between $M_{ab}$ and $M_c$,  but above the metamagnetic transition  $M_{ab}$ $\simeq$ $M_c$. We also observe that $M_{ab}(T) / H$, for 30 kOe, 40 kOe and 55 kOe (shown in the Fig.~\ref{MH}(b)) is similar to $M_{c}(T) / H$. All of these data  indicate a field induced change of magnetic structure.     

\begin{figure}
	\centering
	\includegraphics[width=\columnwidth]{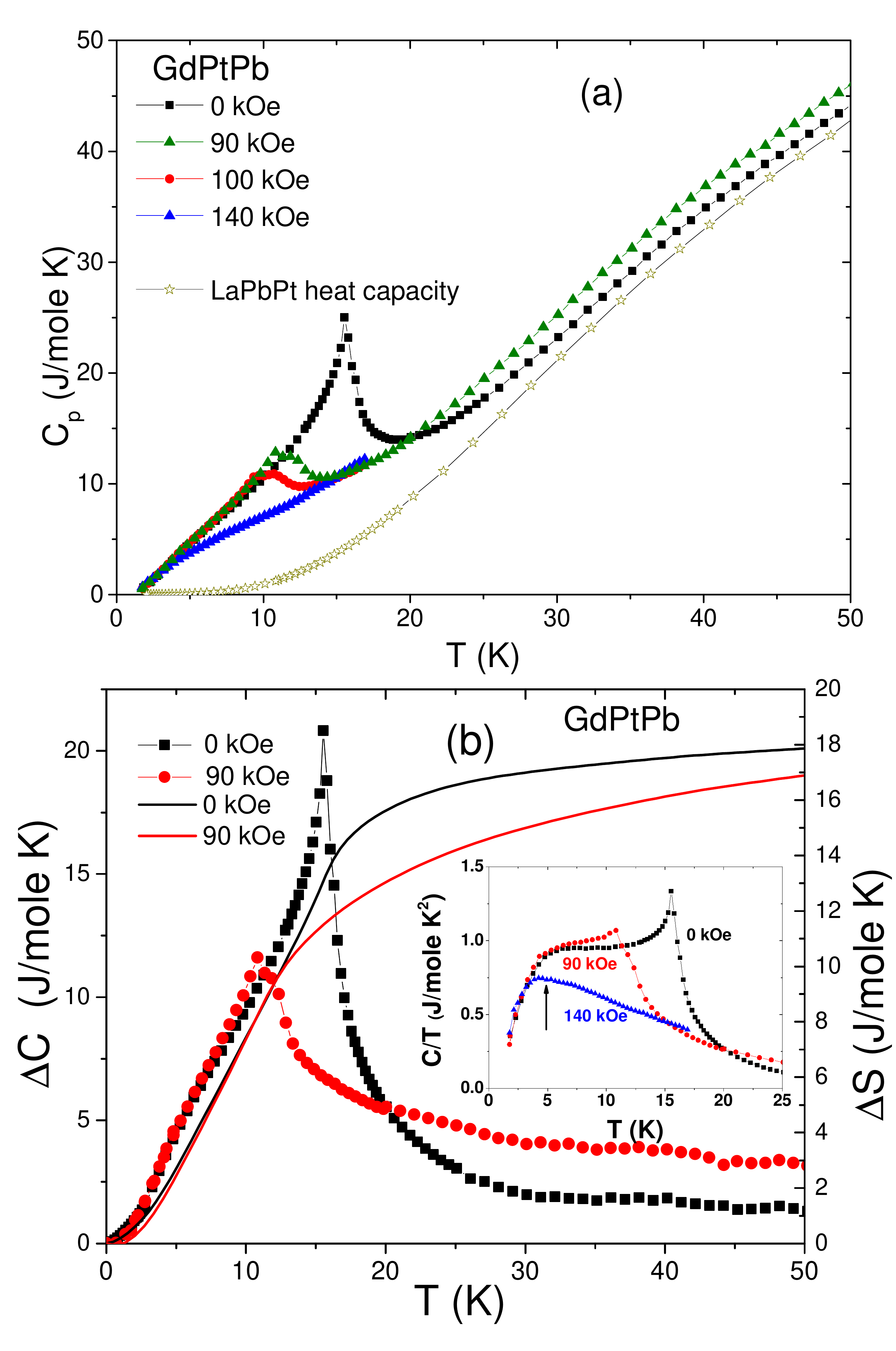}
	\caption{(a)Temperature dependent heat capacity of GdPtPb at zero field at 90 kOe, 100 kOe and 140 kOe field. LaPbPt heat capacity to estimate non-magnetic contribution of heat capacity is also shown. (b) Temperature dependence of magnetic heat capacity ($\Delta C$)  for zero field and 90 kOe field, calculated after a power-law extrapolation of $C$ vs. $T$ down to ($C,T$) = (0,0) and magnetic entropy for the same field values, calculated by integrating $\Delta C/T$ . Inset shows $\Delta C / T$ vs. $T$ for zero field, 90 kOe and 140 kOe field near the low temperature hump (pointed by vertical arrow) without any extrapolation.}\label{Cp}
\end{figure}

Long range AF ordering at $T_N$ = 15.5 K is further confirmed by heat capacity ($C_p$) data. Fig.~\ref{Cp}(a) shows $C_p$ versus $T$ in zero field as well as 90 kOe, 100 kOe and 140 kOe field applied along the $ab$-plane. A sharp, $\lambda$-like, anomaly is observed at $T_N$ in zero field. With increasing field the anomaly shifts to lower temperature and at 140 kOe, no sharp anomaly is observed down to 1.8 K. We have estimated the magnetic contribution of the heat capacity by subtracting LaPbPt heat capacity. 


To calculate the magnetic entropy ($\Delta S$) down to 0 K,  $C$ versus $T$ data is extrapolated to ($C,T$) = (0,0) using a power law fit. Then the magnetic heat capacity ($\Delta C$) is estimated by subtracting LaPbPt heat capacity from extrapolated $C$ and finally, $\Delta C/T$ is integrated over the range between 0 -50 K  [shown in Fig.~\ref{Cp} (b)].  In applied magnetic field, we observe a shift of the sharp anomaly in $\Delta C$ at $T_N$ to lower temperature as well as a shifting of some residual entropy to higher temperature, the later being evident from the tail in $\Delta C$ above $T_N$ for 90 kOe. 

 For zero field, we observe that just above $T_N$, $\Delta S$ reaches up to 77\% of the theoretically expected value for Gd$^{3+}$ which is $R ln (2J+1)$ = $R ln 8$ = 17.2 J/mole K with $J$= 7/2. The remaining entropy is spread well above $T_N$. This indicates that  there is some amount of short range order or fluctuations present above $T_N$. $\Delta S$ saturates to a value little more than $R ln8$ because we have not considered mass correction of LaPbPt heat capacity to estimate magnetic heat capacity.  For 90 kOe, $\Delta S$ does not saturate and is continuously increasing after a slope change at $T_N$. Only about 48\% of magnetic entropy is recovered  near the magnetic ordering temperature at 90 kOe; the remaining entropy is shifted to higher temperature due to partial polarization of the paramagnetic spins along the direction of the magnetic field. Below $T_N$ a broad hump is observed in $\Delta C / T$ [see inset Fig.~\ref{Cp}(b)], which will be discussed below.


\begin{figure}
	\centering
	\includegraphics[width=\columnwidth]{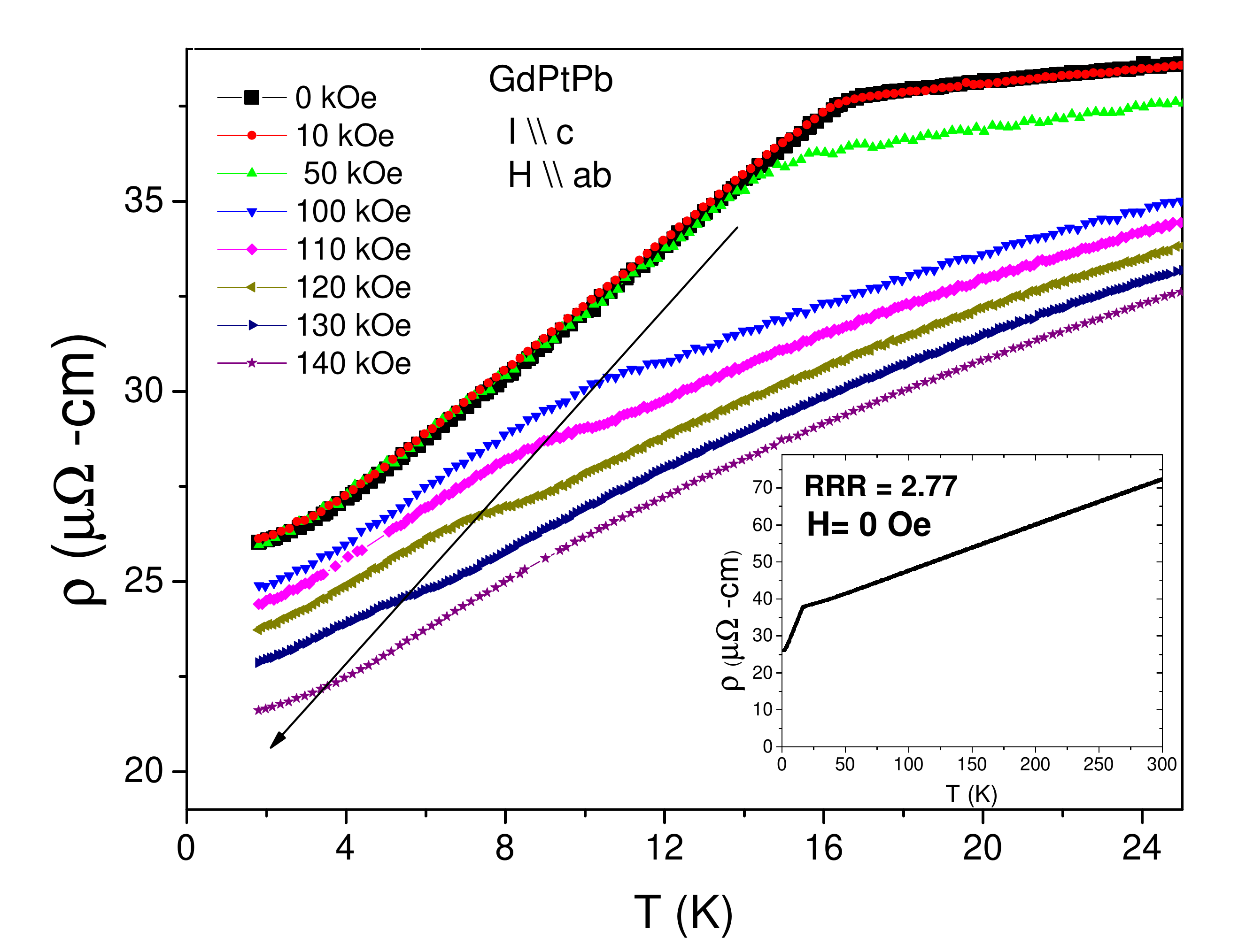}
	\caption{Resistivity ($\rho$) versus $T$ for different field along the $ab$-plane. Inset shows the temperature dependent electrical resistivity ($\rho$) in zero field for current along $c$-axis. }\label{RT}
\end{figure}

\begin{figure}
	\centering
	\includegraphics[width=\columnwidth]{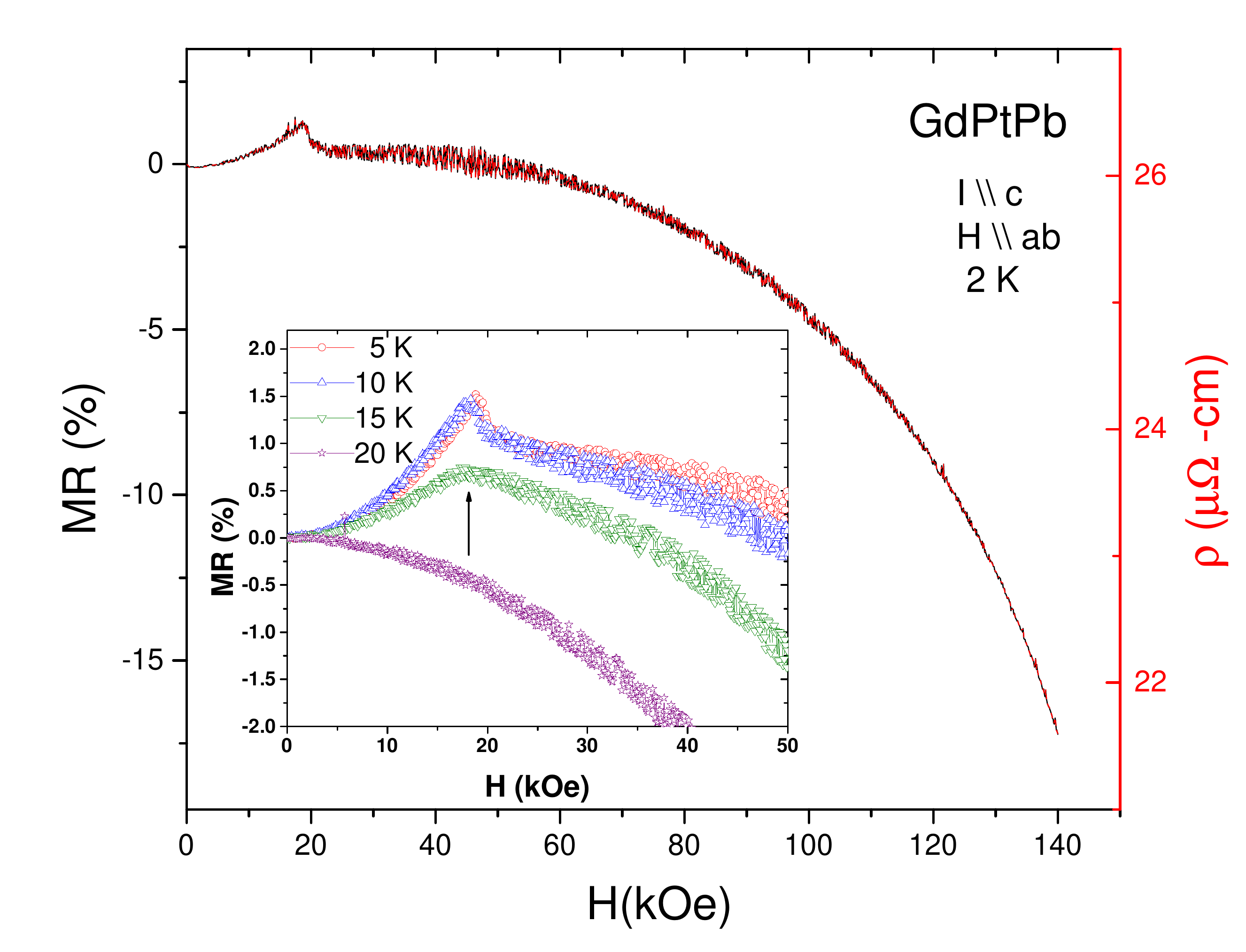}
	\caption{$MR$ versus $H$ (left-axis) $\rho$ versus $H$ (right-axis) at 2K for $H ||ab$. Inset shows magnetoresistance (MR) versus  $H$ near the metamagnetic transition at different temperature, showing the anomaly.}\label{RH}
\end{figure}

Electrical transport measurements on GdPtPb were done by applying current along the $c$-axis and field perpendicular to the $c$-axis. Temperature dependent electrical resistivity, in zero field, is shown in the inset of Fig~\ref{RT}. The room temperature resistivity ($\rho$) is $\sim$ 70 $\mu\Omega$-cm and the residual resistivity ratio [RRR = $\rho$(300K)/$\rho$(1.8K)] is 2.8. Despite the lackluster RRR, we observe a sharp anomaly in $\rho$(T) at $T_N$  due to loss of spin disorder scattering. In an applied field parallel to $ab$-plane, the anomaly shifts to lower temperature, as shown by an arrow in the main panel of Fig~\ref{RT}. At higher fields the anomaly due to loss of spin disorder scattering also weakens and the feature changes. At a 140 kOe  we do not observe any anomaly down to 1.8 K.

Magnetic field dependent electrical transport measurement data are shown in Fig.~\ref{RH}. At 2 K magnetoresistance [MR= [$\rho$(H)-$\rho$(0 kOe)]/$\rho$(0 kOe) x 100] was measured, after an initial increase, followed by a sharp drop at the 20 kOe metamagnetic field, $\rho$ only decreases by 17\% up to 140 kOe ($MR$ = -17\%). In the $MR$ vs. $H$ and $\rho$ vs. $H$ data, measured at 2 K, we observe a sharp kink around 20 kOe. With increasing temperature the sharp kink in MR broadens and vanishes above $T_N$ (see inset). For $T > T_N$, The $MR$ is negative for all the field values measured, consistent with a suppression of spin-disorder scattering in the paramagnetic state associated with Brillouin like polarization of the Gd$^{3+}$ moments. 

\begin{figure}
	\centering
	\includegraphics[width=\columnwidth]{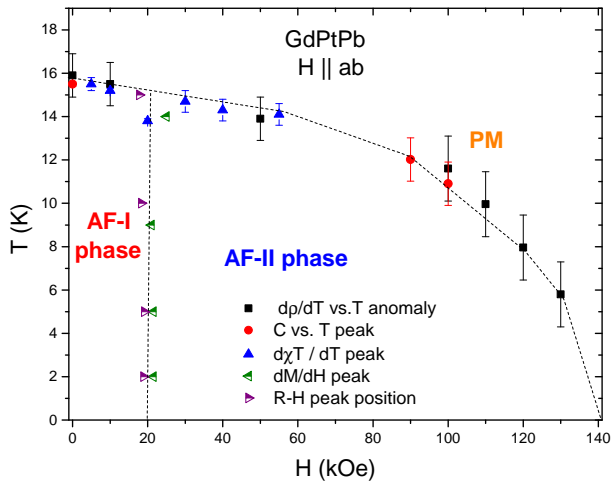}
	\caption{$H$-$T$ phase diagram of GdPtPb for $H||ab$. $T$-axis refers magnetic ordering temperature, $H$-axis refers to magnetic field applied within the $ab$-plane.}\label{phase}
	
\end{figure}
 
\begin{figure}
	\centering
	\includegraphics[width=\columnwidth]{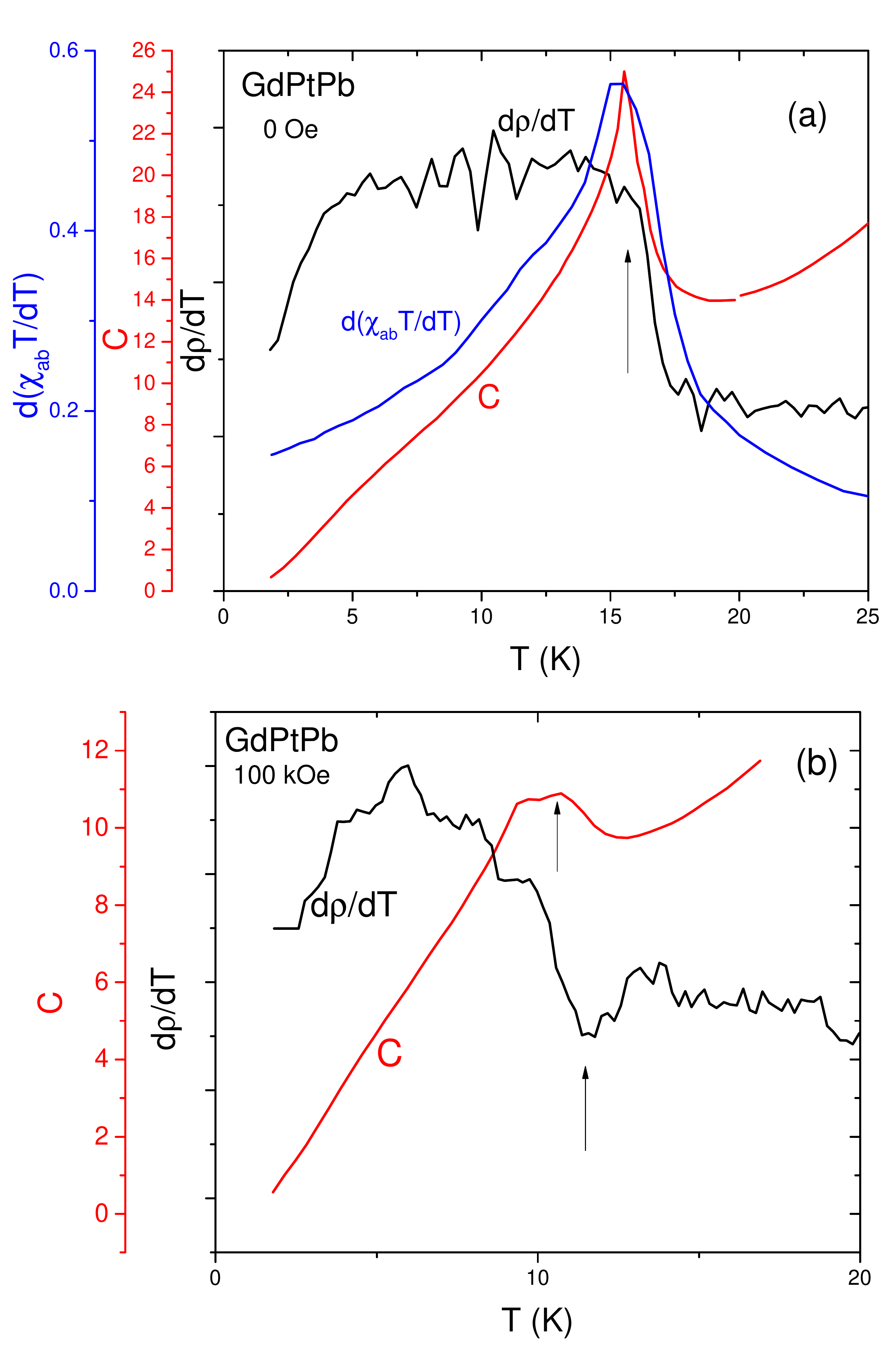}
	\caption{(a) $d(\chi_{ab}T)/dT$, $C$ and $d\rho/dT$ versus $T$ for 0 Oe applied magnetic field, vertical arrow points to $T_N$ (b) $C$ and $d\rho/dT$ versus $T$ for 100 kOe applied magnetic field, vertical arrows point to $T_N$ }\label{fisher}
	
\end{figure}

 Using our magnetization, electrical transport and heat capacity data, we can construct a $H$ - $T$ phase diagram for the magnetically ordered state of GdPtPb for $H || ab$, shown in Fig.~\ref{phase}. For the field parallel to $ab$-plane, the boundary between AF ordered phase and paramagnetic (PM) phase are determined from (1) the $d\rho T/dT$ vs. $T$ anomaly,  which is a jump for 0, 10 and 50 kOe magnetic field (shown for 0 Oe in Fig.\ref{fisher}(a)) and a pronounced minimum for 100, 110, 120 and 130 kOe magnetic field  (shown for 100 kOe in Fig.\ref{fisher}(b)); (2) the peak position in the $d(\chi_{ab} T)/dT$ vs. $T$ (inset Fig.~\ref{MH} ) and (3) peak in $C_p$ vs. $T$ (Fig.~\ref{Cp}). Up to 130 kOe we could track the transition, at 140 kOe, no sharp feature we could associate with a transition is observed down to 1.8 K in resistivity and heat capacity. These data (Fig.\ref{phase}) suggest either a field  induced quantum critical point or a quantum phase transition, most likely to a saturated paramagnetic behavior, near 140 kOe. In addition to the phase boundary of the magnetic order, a change in the magnetic structure around 20-22 kOe is observed. The metamagnetic phase boundary is plotted in the phase digram  from the peak position in the $MR$ vs.$H$ plot (see inset Fig.~\ref{RH}) and peak in $dM/dH$ (inset of Fig.~\ref{MH} (a)).

\section{Discussion and Conclusion}
   From the analysis of the high temperature ($T >$ 50 K) magnetic susceptibility data, we found that $|\Theta_{ab}|$ $\simeq$ $|\Theta_{c}|$  $<<$ $T_N$. Taking $\Theta_{avg}$ $\approx$  -4 K, we find a frustration parameter $|f| = |\Theta_{avg}| / T_N$ $\approx$ 0.25; a value much less than 1.0, suggesting an "anti-frustration" effect. This is contradictory to simple, first order mean field theory (MFT) applied to a single spin sub-lattice (or similar). For a magnetically frustrated material we often observe $|f| >> 1$ which results from the reduced ordering temperature $T_N$ due to competing magnetic exchange interaction in a frustrated lattice.\cite{Ramirez}  We can explain $\Theta$ $<<$ $T_N$ from a more general approach in the MFT.  In the MFT, antiferromagnetism is explained by total magnetism due to interaction between two interpenetrating spin sub-lattices (1 and 2), having spin-up and spin-down. In the first order MFT, the molecular field ($B$) in one sub-lattice is considered to be only proportional to the total magnetization ($M$) in the other sub-lattice, $B = -|\lambda| M$, where $|\lambda|$ is the molecular field constant.\cite{Blundell} In general, the interaction within one sub-lattice can be significantly different from the interaction between two sub-lattices. This leads to more general considerations where we need to consider molecular fields due to the interaction between two sub-lattices (constant given by $|\lambda|$, which is antiferromagnetic) and within a sub-lattice (constant given by $\Gamma$). The molecular fields in two sub-lattices are then given by $B_1 = -\Gamma M_1 - |\lambda| M_2$ and $B_2 = -\Gamma M_2 - |\lambda| M_1$.\cite{Blundell1} Now if we consider an equal number of spins, $n/2$, in the two sub-lattices, from the MFT calculations, $T_N = (|\lambda|- \Gamma) C$ and $\Theta = - (|\lambda|+ \Gamma) C$, where $C$ is the Curie constant.\cite{Blundell1} So if $\Gamma \neq 0$ then $|\Theta| \neq T_N$. In our case, $T_N / \Theta_{avg}$ $\approx$ -4 which would suggest that $|\lambda| / \Gamma$ $\approx$ -1.67. If we consider a simple two sub-lattice picture of antiferromagnetism for GdPtPb, we can assume $J_1$ and $J_2$ are the nearest neighbor and  the second nearest neighbor exchange interactions proportional to $\Gamma$ and $|\lambda|$ respectively within the $ab$-plane (we are assuming exchange interaction along $c$-axis will be similar for the two sub-lattices). So we get $J_2/J_1$ $\approx$ -1.67 and they have opposite sign. Since $|\lambda|$ is antiferromagnetic and $J_2 > 0$, we get $J_1 < 0$  and ferromagnetic. In our case,  $|J_2| > |J_1|$, hence average exchange interaction which is proportional to $\Theta_{avg}$ is small and antiferromagnetic type. Similar analysis was done in EuRh$_2$As$_2$ by Singh et. al.\cite{Singh}. Hence MFT analysis suggests that for GdPtPb, the antiferromagnetic  $J_2$ is greater than the ferromagnetic $J_1$. This is possible for RKKY-type exchange interaction which follows a oscillatory decay function in space.

The low-$T$ magnetic susceptibility, below the metamagnetic transition field (20 kOe) gives $\chi_{ab}$($T$ $\to$ 0)/ $\chi_{ab}$($T_N$) $\approx$  1/2.  This strongly suggest a planar non-collinear magnetic structure within the Kagom\'e sub-lattice with the spins being in the $ab$-plane. The non-collinear structure can either be intrinsic or may originate from three domains of collinear spins rotated by 120$^{\circ}$ to each other in the hexagonal $ab$-plane. To determine the exact spin structure and magnetic $\textbf{Q}$-vector, microscopic measurements are underway. We designate this antiferromagnetic phase as AF-I in the phase diagram (see Fig.~\ref{phase}). Above the metamagnetic transition field,  $\chi_{ab}$($T$ $\to$ 0)/ $\chi_{ab}$($T_N$) $\approx$  1 and $\chi_{ab}$ = $\chi_c$. We designate this as AF-II phase (see Fig.~\ref{phase}).

The broad hump, observed in $\Delta C / T$ below $T_N$  [see inset Fig.~\ref{Cp}(b)] is also weakly visible in the $\Delta C$. In some cases such low-$T$ hump in heat capacity originates from partial disorder of the spins due to structural defects,\cite{Manni} and vanishes with the better ordering in the single crystalline material.\cite{Freund}. We have not observed any structural disorder in GdPtPb. In addition to that, we observed that the position of the that hump in $\Delta C / T$ does not shift or significantly broaden with the increasing magnetic field, a stark contrast to the structural disorder scenario.\cite{SG} Hence the structural disorder is not the reason behind the low-$T$ broad hump in $\Delta C / T$.  For 140 kOe magnetic field, when the antiferromagnetic ordering is suppressed below 1.8 K , the broad hump in $\Delta C/T$ around  4 K still survives.  This strongly suggests that this feature is not related to magnetic ordering. Such a broad hump in $\Delta C$ below the $\lambda$-like anomaly at $T_N$ is observed in some other Gd-based systems like GdBiPt~\cite{gdbipt}, GdCu$_2$Si$_2$~\cite{gdcusi} and GdFe$_2$Ge$_2$ ~\cite{gdfege}. A very similar feature is observed in the calculated magnetic heat capacity from the MFT\cite{Johnston2} where the broad hump increases with increasing value of $S$ ($J$) and at the classical limit of spin $S=10$, the $\Delta C$ does not go to zero rather saturate to a finite value.\cite{bamnas} This indicates that this Schottky-like anomaly appears due to Zeeman-splitting  of the $2J+1$ multiplet under the internal magnetic field. This becomes experimentally distinguishable in case of only a Gd-based compound where whole 2$J$+1 multiplet participates in the magnetism instead of the ground state doublet and is most likely origin of the feature we observe in GdPtPb.           
  
 In Summary, the search for Gd-based frustrated AF in the ZrNiAl-type distorted Kagom\'e structure lead us to the discovery of antiferromagnet, GdPtPb, in single crystalline form which magnetically orders with a planar non-collinear magnetic structure below 15.5 K and undergoes a field induced change in the magnetic structure around 20 kOe.  
  
 We conclude that GdPtPb can serve as an example of mean field non-collinear AF on hexagonal lattice with a distorted Kagom\'e magnetic sub-lattice.    
 
 SM thanks David C. Johnston and Valentin Taufour for very useful discussion. SM was funded by the Gordon and Betty Moore Foundations  EPiQS Initiative through Grant GBMF4411. This work was supported by the US Department of Energy, Office of Science, Basic Energy Sciences, Materials Science and Engineering Division. Ames Laboratory is  operated for the US Department of Energy by Iowa State University under contract No. DE-AC02-07CH11358.


\end{document}